\documentclass[aps,showpacs,amsmath,amssymb,nofootinbib,preprintnumbers]{revtex4}

\usepackage{graphicx}
\usepackage{color}

\def \be  {\begin{equation}}
\def \ee  {\end{equation}}
\def \bea {\begin{eqnarray}}
\def \eea {\end{eqnarray}}

\begin{document}

\preprint{ECTP-2010-02}

\title{Bulk and Shear Viscosity in Hagedorn Fluid}
\author{A.~Tawfik}
\email{drtawfik@mti.edu.eg}
\affiliation{Egyptian Center for Theoretical Physics (ECTP), MTI University,
 Cairo-Egypt}
 \author{M.~Wahba}
\affiliation{Egyptian Center for Theoretical Physics (ECTP), MTI University,
 Cairo-Egypt}

\begin{abstract}
Assuming that the Hagedorn fluid composed of known particles and resonances with masses $m<2\,$GeV obeys the {\it first-order} theory (Eckart) of relativistic fluid, we discuss the transport properties of QCD confined phase. Based on the relativistic kinetic theory formulated under the relaxation time approximation, expressions for bulk and shear viscosity in thermal medium of hadron resonances are derived. The relaxation time in the Hagedorn dynamical fluid exclusively takes into account the decay and eventually van der Waals processes. We comment on the {\it in-medium} thermal effects on bulk and shear viscosity and averaged relaxation time with and without the excluded-volume approach. As an application of these results, we suggest the dynamics of heavy-ion collisions, non-equilibrium thermodynamics and the cosmological models, which require thermo-- and hydro--dynamics equations of state.
\end{abstract}
\date{\today}

%

\maketitle

\section{Introduction}\label{2sec:intro}
Since the discovery of {\it new matter} at the relativistic heavy-ion
collider (RHIC) \cite{larry}, an immense number of experimental and theoretical works have been
invested to explore the properties of {\it non-confined} strongly
coupled matter, the quark-gluon plasma (QGP). Characterizing QCD matter through
transport and collective behavior and the equation of state is the
ultimate goal. Also, the {\it confined} QCD matter has been subject of various
papers \cite{V-hadron1,V-hadron2,redl1,etas2}.  

Shear viscosity $\eta$ characterizes the elliptic flow and is directly proportional to energy density $\epsilon$
and inversely to scattering cross section $\sigma$ and simultaneously 
reflects how particles interact and collectively
move in many particle medium. Strongly interacting matter, like Hagedorn gas
\cite{hagdr1} is conjectured to have smaller
$\eta$ than the weakly interacting one. Using perturbative and non-perturbative
methods, $\eta$, (and normalized to entropy $s$)
has been estimated for {\it non-confined} and {\it confined} QCD matter,
respectively. 

At temperature $T>>T_c$, bulk viscosity $\xi$ has been studied using
perturbative QCD \cite{perturQCD1} and found to be negligible comparing to
$\eta$ and regarding to the collective evolution of the many body
system, as QCD turns to be conformal invariance. Recent lattice QCD
simulations~\cite{mueller2} show that bulk 
viscosity $\xi$ is not negligible near $T_c$. Its rapid increase at $T_c$ is
apparently associated with a fast growth of the trace anomaly,
$(\epsilon-3p)/T^4$, of the energy-momentum tensor $T^{\mu\nu}$
\cite{karsch1}. Below $T_c$, various hadron scales likely provide the
conformal invariance with {\it bad} symmetries (QCD conformal anomalous). It
is therefore natural to expect that $\xi$ at this energy scale is not
negligible \cite{QCD-conformal1}. 

There have been many attempts to compute $\xi$ and $\eta$ in the {\it confined}
QCD matter using effective models for the hadron resonances \cite{eta-hadrons1,etas2,etas2b,eta-hadrons2,eta-hadrons3a,eta-hadrons3b,eta-hadrons4}. In
this letter, we discuss the bulk and shear viscosity of the Hagedorn dynamical fluid using the relativistic kinetic theory in the 
relaxation time approximation \cite{V-hadron1,V-hadron2,redl1} and the explicit 
implementation of the hadronic mass spectrum and excluded--volume approach.   

Disregarding all interactions but decay and repulsion, the thermal change of the relaxation time in the Hagedorn fluid has been studied, which complete a set of thermo \cite{tawf2} and hydrodynamic equations of state needed to characterize the evolution equation in early universe, for instance.  

\section{Bulk and Shear Viscosity}\label{2sec:model}

The relativistic kinetic theory gives the transport equations for classic and colored particles in a non--Abelian external field. In this letter, we
discuss the transport coefficients for {\it confined} QCD matter
composed of known particles and resonances. We apply the relaxation time approximation of
Boltzmann equation and take into consideration the hadronic density states in
{\it confined} QCD. The transport equations describe the evolution of the phase space
distribution function of the particles of interest. The transport properties
are defined as the coefficients of spatial component of the difference between
the energy-momentum tensors out of and at equilibrium corresponding to the Lagrangian density.  

We model the Hagedorn fluid of QCD {\it confined} phases (fermion and boson resonances) as a non–interacting gas of hadron resonances. To do so we sum over all Fermi and Bose resonances. The main motivation of doing this is that it refers to all relevant degrees of freedom of the {\it confined} strongly interacting matter. It implicitly includes the interactions that likely result in resonance formation \cite{hresons}. Natural units $c=\hbar=k=1$ are applied here. The Hagedorn mass spectrum $\rho(m)$ implies growth of the hadron mass spectrum with increasing the resonance masses. 
\begin{eqnarray}\label{eq:rhom}
\rho(m) &=& c\left(m_0^2+m^2\right)^{k/4} \exp(m/T_H),
\end{eqnarray}
with $k=-5$, $c=0.5\,$GeV$^{3/2}$, $m_0=0.5\,$GeV and $T_H=0.195\,$GeV. 
 
This model provides a quite satisfactory description of particle production in heavy-ion collisions \cite{tawf1,tawf2,reff1}. The repulsive interactions likely soften the $T$--dependence of the thermodynamic quantities. The excluded--volume approach \cite{extndd} is used to implement the effects of repulsive interactions (van der Waals) by assuming the energy normalized by $4{\cal B}$ equals the  excluded--volume and the intensive quantity $T_{pt}$ in the point--type particle approach (and the other thermodynamic quantities \cite{extndd2}) have to be {\it corrected} as follows.
\begin{eqnarray}
T &=& \frac{T_{pt}}{1-\frac{P_{pt}(T_{pt})}{4{\cal B}}}, \hspace*{2cm}\left(p(T) = \frac{p_{pt}(T_{pt})}{1-\frac{P_{pt}(T_{pt})}{4{\cal B}}}\right)
\end{eqnarray}
where ${\cal B}^{1/4}=0.34\,$GeV stands for the MIT bag constant.

In spherical polar coordinates, the energy--momentum tensor of a single particle with $p$-- and $T$--independent mass $m$ is defined as  
\begin{eqnarray}
T^{\mu\nu}_1 &=& \frac{g}{2\pi^2} \rho(m)\int p^2\, dp \;
\frac{p^{\mu}p^{\nu}}{\varepsilon}\;  n(p,T),   
\end{eqnarray}
where $p^{\mu}=(\varepsilon,\vec{p})$ is momentum four-vector and $g$ is degeneracy factor of the hadron resonances. The single particle energy is given by the dispersion relation 
$\varepsilon=(\vec{p}\,^2+m^2)^{1/2}$. 

With the above assumptions on Hagedorn viscous fluid, the overall energy--momentum tensor can be calculated as a sum over energy-momentum tensors $T^{\mu\nu}_1$ of all hadrons resonances \footnote[1]{It reflects the algebraic properties, here the addition, of the energy-momentum tensor},
\begin{eqnarray}
T^{\mu\nu} & =& \sum_i T^{\mu\nu}_i 
\end{eqnarray}
In momentum phase space and assuming that the system is in a state with vanishing chemical potential and near equilibrium, the distribution function $n(p,T)$ reads
\begin{eqnarray}\label{eq:noTp}
n(p,T) & =& \frac{1}{\exp\left(\frac{\varepsilon - \vec{p}\cdot\vec{u}}{T}\right)\pm1}, 
\end{eqnarray}
where $\pm$ stands for fermion and boson statistics, respectively. The local flow velocity $\vec{u}$ is compatible with the Eckart fluid \cite{eckrt}, implying that $T^{\mu\nu}u_{\mu}u_{\nu}=\varepsilon$. It is obvious that $n(p,T)$ satisfies the kinetic theory \cite{V-hadron2} and second law of thermodynamics. The solution of kinetic equation is obtainable by deviating the distribution function from its local equilibrium.

The deviation of energy-momentum tensor from its local equilibrium is corresponding to the difference between the distribution function near and at equilibrium, $\delta n=n-n_0$. The latter can be determined by relaxation time approximation with vanishing external and self-consistent forces \cite{V-hadron2,relaxx}
\begin{eqnarray}
\delta n(p,T) &=& - \tau \frac{p^{\mu} }{\vec{p}\cdot\vec{u}} \partial_{\mu} n_0(p,T)
\end{eqnarray}

\begin{figure}
\centering
\includegraphics[width=8.5cm]{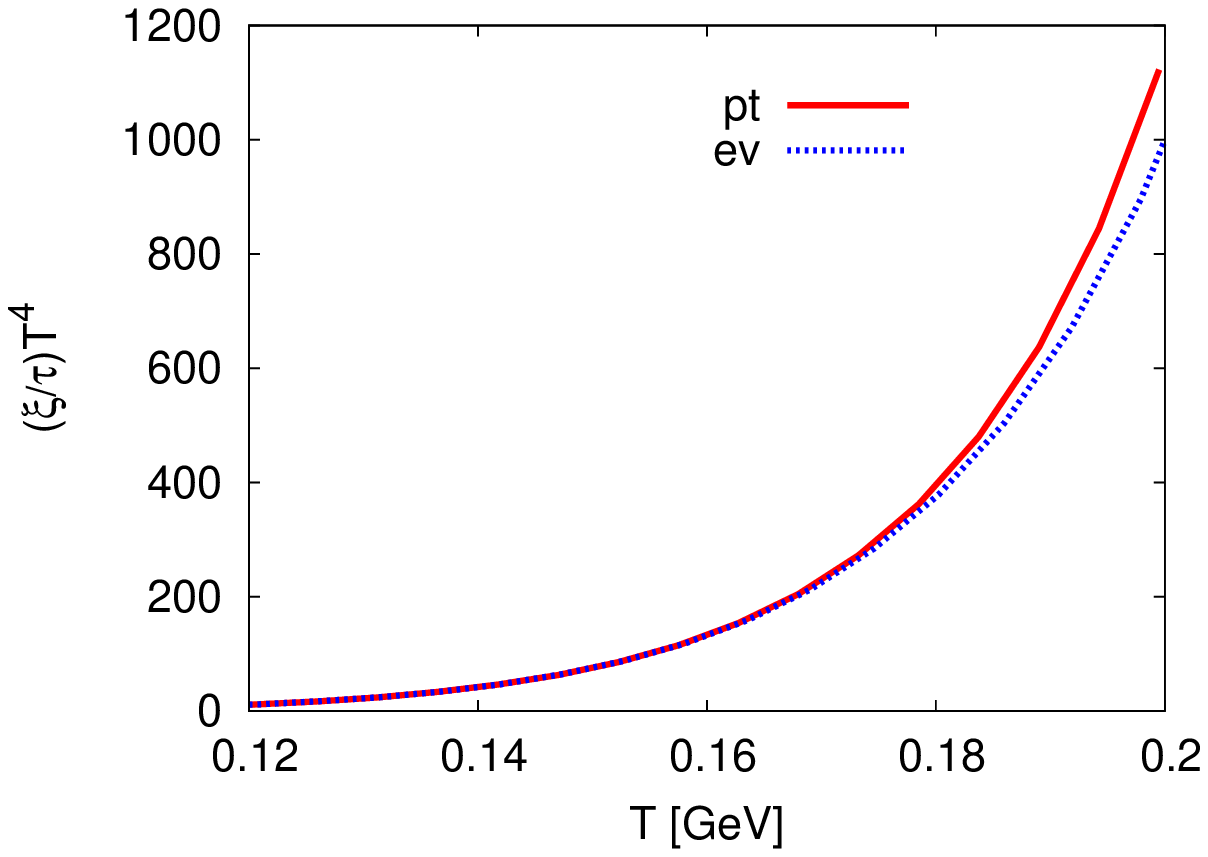}
\includegraphics[width=8.5cm]{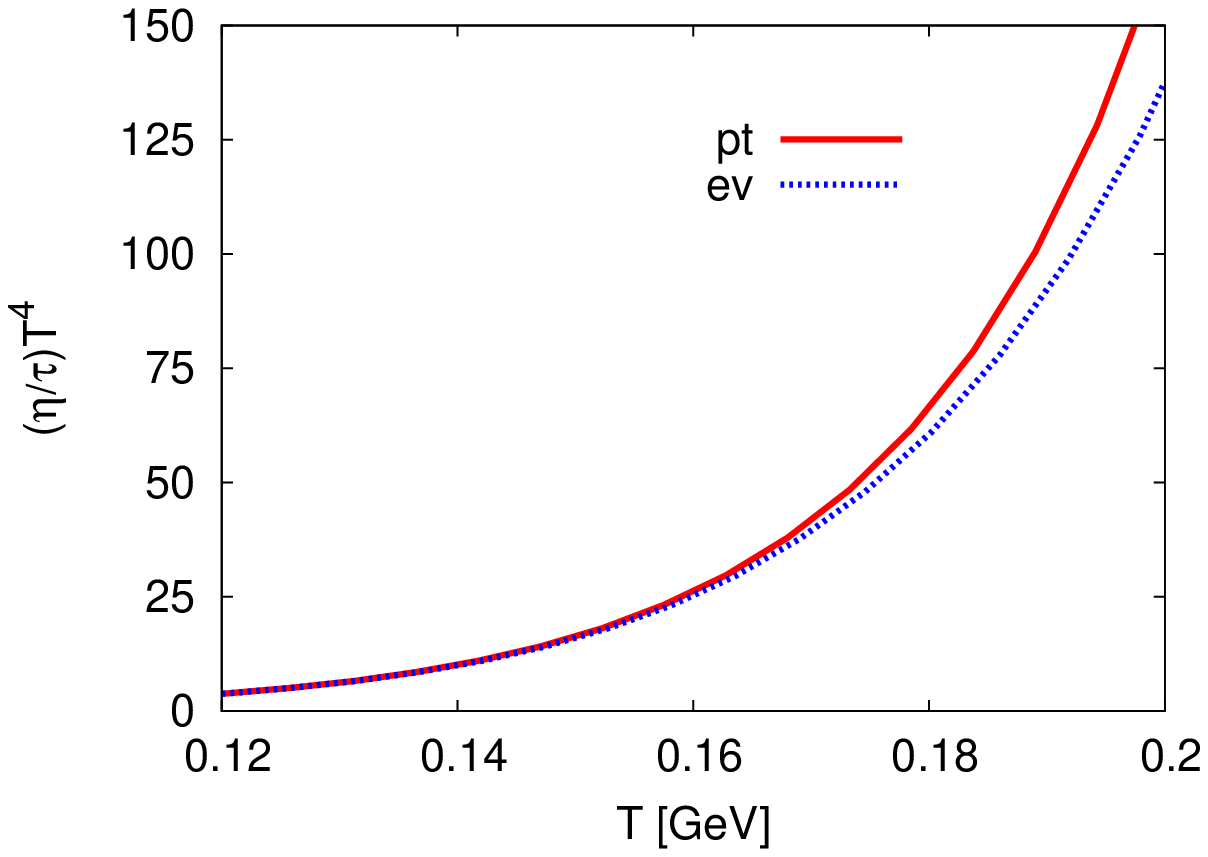}
\caption{Bulk (left panel) and shear (right panel) viscosity coefficients of Hagedorn fluid are depicted as function of the heat bath temperature. The effects of excluded-volume approach are illustrated (dashed lines).}
\label{2fig1}
\end{figure}

Then the difference between near and equilibrium energy--momentum tensor reads
\begin{eqnarray}
\delta T^{\mu\nu}_1 &=& -\frac{g}{2\pi^2} \rho(m)\int_{0}^{\infty} p^2\, dp \;
\frac{p^{\mu}p^{\nu}}{\varepsilon^2}\; \tau\; p^{\alpha}\partial_{\alpha}\; n_0(p,T).   
\end{eqnarray}
Using the symmetric projection tensor $h_{\alpha\beta}$ \cite{maartns}, the components of the derivative $\partial_{\alpha}$ can be split to parallel and orthogonal to $u^{\mu}$. $h_{\alpha\beta}$ generates a 3-matric and projects each point into the instantaneous rest space of the fluid.  
\begin{eqnarray}
\partial_{\mu} &=& D u^{\mu} + \nabla_{\mu},
\end{eqnarray}
where $D=u^{\alpha} \partial_{\alpha}=(\partial_{t},0)$ gives the temporal derivative and $\nabla_{\mu}=\partial_{\mu}-u_{\mu} D =(0,\partial_{i})$ give the spacial derivative \cite{V-hadron2}. Such an splitting has to guarantee the conservation of equilibrium energy--momentum tensor; $\partial_{\mu} T^{\mu\nu}=0$ and fulfill the laws of thermodynamics at equilibrium \cite{V-hadron2,maartns}. In ref \cite{V-hadron1}, the non-equilibrium $n(p,T)$ has been decomposed using the relaxation time approach $n=n_0+\tau n_1+\cdots$. Alternatively, as $n(p,T)$ embeds the 1$^{st}$-rank tensor $u$, $\delta T^{\mu\nu}_1$ can be decomposed into $u$ \cite{V-hadron2} in order to deduce its spatial components.
\begin{eqnarray}\label{eq:decomp}
\partial_k u^l &=& \frac{1}{2}\left(\partial_k u^l + \partial_l u^k - \frac{2}{3}\delta_{kl}\partial_i u^i\right) + \frac{1}{3}\delta_{kl}\partial_i u^i \equiv \frac{1}{2} W_{kl} + \frac{1}{3}\delta_{kl}\partial_i u^i.
\end{eqnarray}
Applying the equation of hydrodynamics, then the deviation from equilibrium 
\begin{eqnarray}
\delta T^{\mu\nu}_1 = \frac{g}{2\pi^2}\rho(m)\tau \int_{0}^{\infty} 
\frac{p^{\mu}p^{\nu}}{T} n_0(1+n_0) \left[\vec{p}\cdot\vec{u} c_s^2 \nabla_{\alpha} u^{\alpha} + p^{\alpha}\left(\frac{\nabla_{\alpha} p}{\epsilon+p} - \frac{\nabla_{\alpha} T}{T}\right)+\frac{p^{\alpha}p^{\beta}}{\vec{p}\cdot\vec{u}} \nabla_{\alpha} u_{\beta}\right] p^2 dp,
\end{eqnarray}
can be re-written as 
\begin{eqnarray} \label{finaldT}
\delta T^{i j}_1 &=& \frac{g}{2\pi^2}\frac{\tau}{T}\rho(m)\int_{0}^{\infty} 
p^{i}p^{j} n_0(1+n_0) \left[\left(\varepsilon c_s^2-\frac{\vec{p}\, ^2}{3\varepsilon}\right)\partial_i u^i - 
\frac{p^kp^l}{2\varepsilon} W_{kl}\right] p^2 dp,
\end{eqnarray}
where $c_s^2=\partial p/\partial \epsilon$ is the speed of sound in this viscous fluid. The bulk $\xi$ and shear $\eta$ viscosity can be deduced from Eq. \ref{finaldT} by comparing it with  
\begin{eqnarray} \label{eq:dltT}
\delta T^{i j}_1 &=& -\xi \delta_{ij} \partial_k u^k - \eta W_{ij}.
\end{eqnarray}
To find shear viscosity $\eta$, we put $i\neq j$ in Eqs.  (\ref{eq:decomp}) and (\ref{eq:dltT}). To find bulk viscosity $\xi$, we substitute $i$ with $j$ and $T^{\mu\nu}_0$ with $3P$. The subscript $0$ (as in the distribution function $n$) refers to the equilibrium state. Although we keep the gradients of velocity, we put $\vec{u}=0$ in the final expressions. 
The intensive quantities $\eta$ and $\xi$ of Hagedorn fluid \footnote[2]{As we assume a vanishing chemical potential, the heat conductivity vanishes as well.}  in the comoving frame read
\begin{eqnarray}
\xi(T) &=& \frac{g}{2\pi^2} \frac{\tau}{T} \sum_i\rho(m_i)\int_{0}^{\infty} n_0(1+n_0)\left(c_s^2 \varepsilon_i^2 - \frac{1}{3}\vec{p}\, ^2 \right)^2 p^2 dp\\
\eta(T) &=& \frac{g}{30\pi^2} \frac{\tau}{T} \sum_i\rho(m_i)\int_{0}^{\infty} n_0(1+n_0) \frac{\vec{p}\, ^4}{\varepsilon_i^2} p^2 dp.
\end{eqnarray}
The $T$--dependence of dimensionless ratios $\xi T^4/\tau$ and $\eta T^4/\tau$ is depicted in Fig. \ref{2fig1}. With increasing $T$, bulk and shear viscosity increase, significantly. We note that $\xi$ seems to be about one order of magnitude larger than $\eta$. Left panel of Fig. \ref{2fig2} illustrates such a comparison. At low $T$, $\eta$ starts with larger values than $\xi$'s. But with increasing $T$, $\xi$ gets larger \cite{xiovereta1}. 

The ratio $\xi/\eta$ has been related to the speed of sound $c_s^2$ in gas of massless pions. Apparently there are essential differences between this system and the one of Hagedorn fluid. According to \cite{etaxiratio1}, the ratio of $\xi/\eta$ in $N=2^*$ plasma is conjectured to remain finite across the second--order phase transition. This behavior seems to be illustrated in Fig. \ref{2fig1}. In the Hagedorn fluid, the system is assumed to be drifted away from equilibrium and it should relax after a characteristic time $\tau$. Should we implement a phase transition in the Hagedorn fluid, then $\tau\propto \xi^{z}$, where $z$ is the critical exponents, likely diverges near $T_c$.

\section{Relaxation Time}

The relaxation time depends on the relative cross section as
\begin{eqnarray}
\tau(T) &=& \frac{1}{n_f(T)\langle v(T)\sigma(T)\rangle},
\end{eqnarray}
where $v(T)$ and $n_f(T)$ is the relative velocity of two particles in case of binary collision and the density of each of the two species, respectively. The thermal-averaged transport rate or cross section is $\langle v(T)\sigma(T)\rangle$. The transport equation of single-particle distribution function in the momentum space, $n(r,p,t)$ \cite{GSt}, 
\begin{eqnarray}\label{BUU}
\frac{\partial}{\partial t} n +\vec{v}\cdot\vec{\nabla}_{r} - \vec{\nabla}_{r}U\cdot \vec{\nabla}_{p} n &=& -\int\frac{d^3p_2 d^3p_1^{\prime}, d^3p_2^{\prime}}{(2\pi)^6} \sigma v \nonumber \\ && \hspace*{4mm} \left[nn_2(1-n_1^{\prime})(1-n_2^{\prime}) - n_1^{\prime}n_2^{\prime}(1-n)(1-n_2)\right]   \\ &&\hspace*{4mm}    
	\delta^4(p+p_2-p_1^{\prime}-p_2^{\prime}). \nonumber
\end{eqnarray}
First line in r.h.s. of Eq. (\ref{BUU}) gives the Boltzmann collision term. The second line adds the Uehling--Uhlenbeck factors. The third line accounts for the Pauli-blocking of the final states. The total derivative of $n$ is given by the collision integral. To solve Eq. (\ref{BUU}), several gradients must be take into account. Real and imaginary parts of the G-Matrix \cite{gmat} are taken to describe the potential of nuclear interactions and the cross section of the binary interaction, respectively. In-medium effects in final (Pauli blocking) and also in intermediate states have to be taken into consideration.
In Ref. \cite{ref:tau1}, the kinetic Boltzmann--Uehling--Uhlenbeck equations for pure nucleon system have been analyzed and the relaxation time in non-relativistic approximation has been deduced as
\begin{eqnarray}\label{eq:NN}
\tau(T) & \approx & \frac{850}{T^2}\left(\frac{n}{n_0}\right)^{1/3} \left[1+0.04 T\frac{n}{n_0}\right] + \frac{38}{\sqrt{T}(1+\frac{160}{T^2})} \frac{n_0}{n}
\end{eqnarray}
where $n$ is the baryon density and $n_0$ is the nuclear saturation density $n_0\approx 0.145 $fm$^{-3}$. 

When fitting the decay widths $\Gamma_i$ of $i$ hadron resonances, then the decay relaxation times $\tau_i$ in GeV$^{-1}$ read $\tau_i\equiv\Gamma_i^{-1}=\left(0.151 m_i -0.058\right)^{-1}$ \cite{eta-hadrons2,extndd2,eta-hadrons5}. As the resonance mass $m$ is conjectured to remain constant in thermal and dense medium, this linear fit apparently implies that $\tau$ remains unchanged as well.

In the Hagedorn fluid, where the inter-particle collisions as in Eq. (\ref{eq:NN}) are minimized, we are left with specific processes to estimate $\tau$ (decay and repulsion for instance). Formation from {\it free space} vacuum and decay to stable resonances; $P_1+P_2\leftrightarrow P_3$ \cite{newRafls} are examples. The constrains on this process have been discussed in \cite{newRafls}. 

In rest of frame of the particle $P_3$ boosting from the laboratory frame, the kinetic equation for the time evolution of the number density $n_3(T)$ reads
\begin{eqnarray}\label{eq:noT}
\frac{d}{d t} n_3(T) = \frac{d}{dVdt}\left(W_{12\rightarrow3} - W_{3\rightarrow12}\right)
\end{eqnarray}
The backward (inverse) direction is also valid. Note that $n(T)$, Eq. (\ref{eq:noT}), and $n(p,T)$, Eq. (\ref{eq:noTp}), are related with each other via $n(T)=N(T)/V=g/(2\pi^2)\sum_i\rho(m_i)\int p^2 dp\, n_i(p,T)$ and therefore $n(p,T)$ is a Lorentz scalar whereas $n(T)$ not. The thermal decay and production rate $dW/dVdt$ have been discussed in \cite{newRafls}. 
In Boltzmann limit and assuming that the repulsive interaction does not contribute meaningfully to the overall relaxation time, the decay time in rest frame is given in textbooks. 
\begin{eqnarray}\label{eq:finalTau}
\tau &=& \frac{8\pi m_3^2 g_3 I}{p\sum_{spin} |\langle\vec{p},-\vec{p}|{\cal M}|m_3\rangle|^2} \left\langle\frac{\varepsilon_3}{m_3}\right\rangle,
\end{eqnarray}
where $p$ is the pressure. $I$ is a step functions for particle distinguishaility; $I=2$ for indistinguishable and $I=1$ for distinguishable particles. ${\cal M}$ is the hadronic reaction matrix. 

\begin{figure}
\centering
\includegraphics[width=8.5cm]{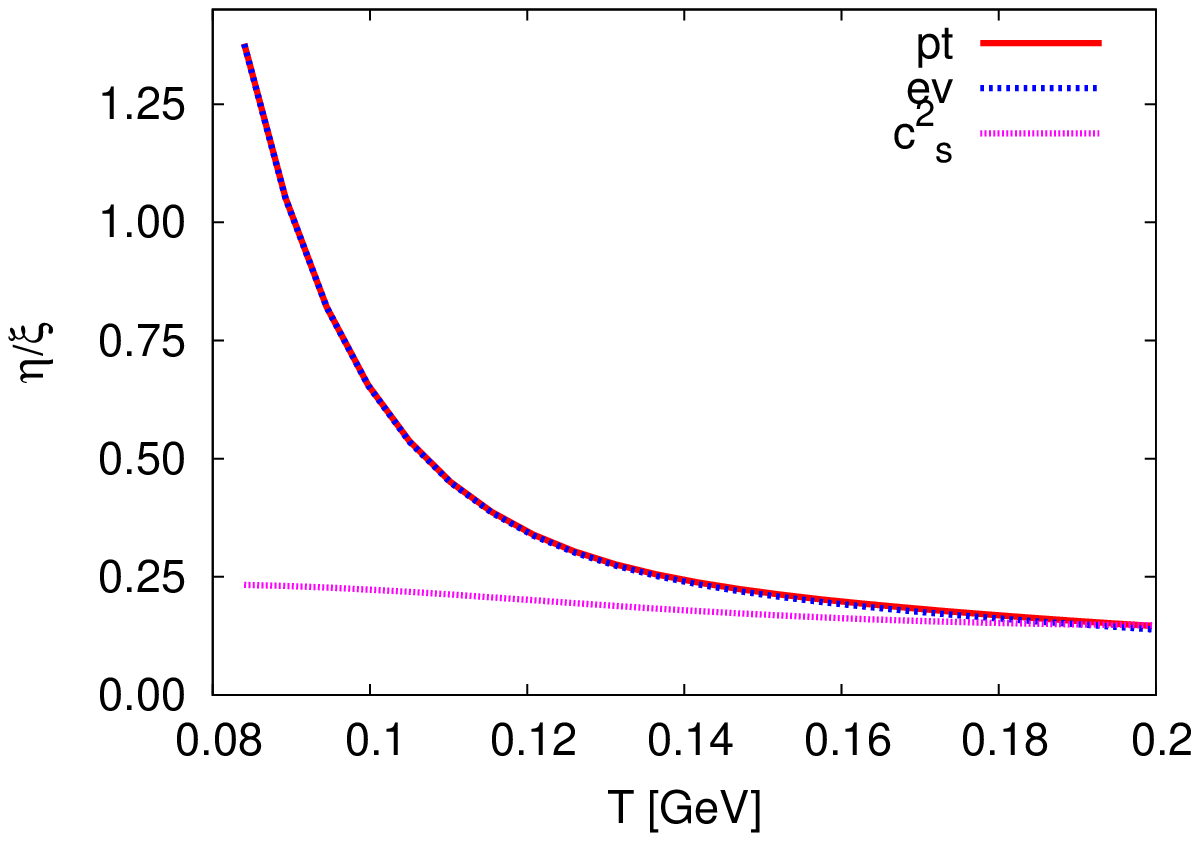}
\includegraphics[width=8.5cm]{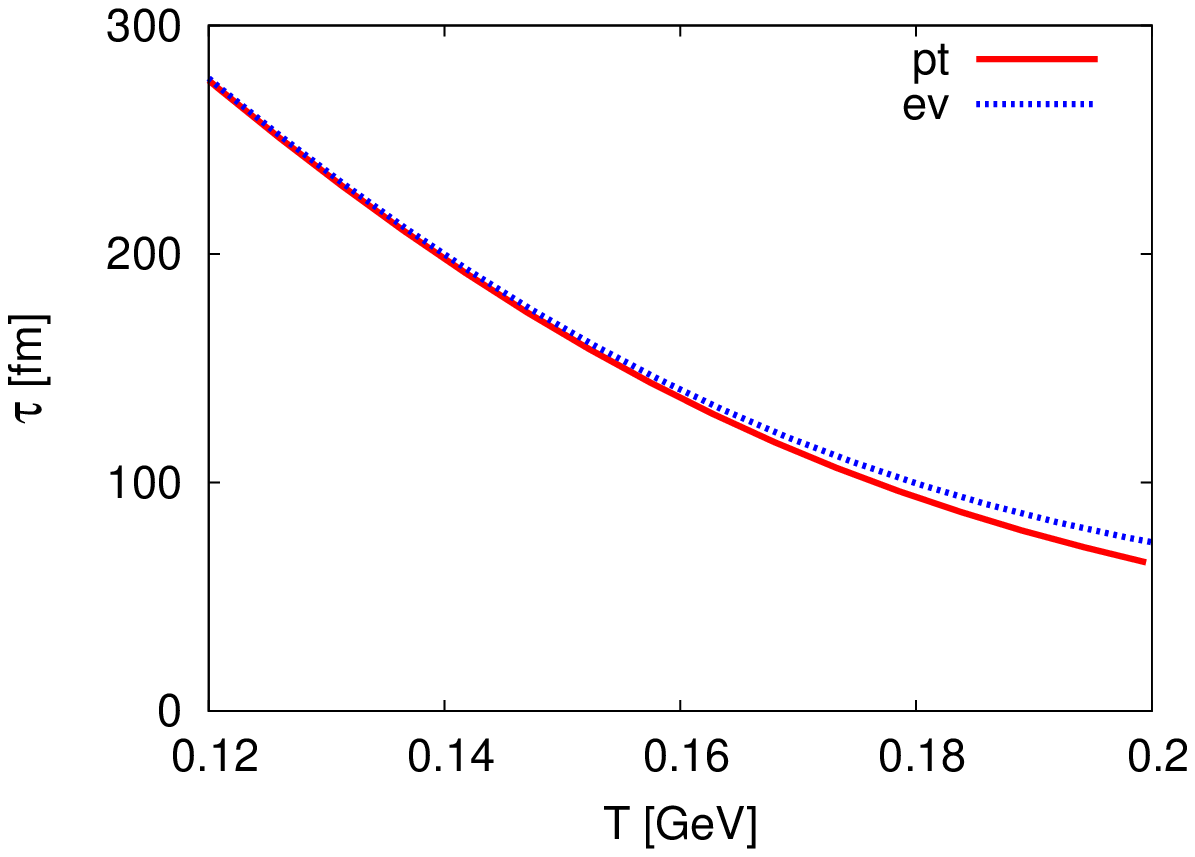}
\caption{Left panel: ratio $\eta/\xi$ vs. $T$. Right panel: thermal change of the relaxation time in Hagedorn fluid. The point-type and excluded-volume approaches are compared. Dashed-dotted line gives the speed of sound.}
\label{2fig2}
\end{figure}

In right panel of Fig. \ref{2fig2}, the relaxation time in fm, Eq. (\ref{eq:finalTau}), is given as function of $T$ in GeV. We note that increasing the temperature $T$ leads to reducing the relaxation time $\tau$. It might mean that the decay processes get faster when $T$ increases. Near $T_c$, the effect of excluded--volume approach is considerable. 

As an application of these results, we mention the cosmological {\it viscous} models \cite{tawf3}, which require a complete set of thermo \cite{tawf2} and hydrodynamic equations of state in order to solve the evolution equation in early universe and study the nucleosynthesis.

\section{Ratio of shear viscosity over entropy density}

\begin{figure}
\centering
\includegraphics[width=14cm]{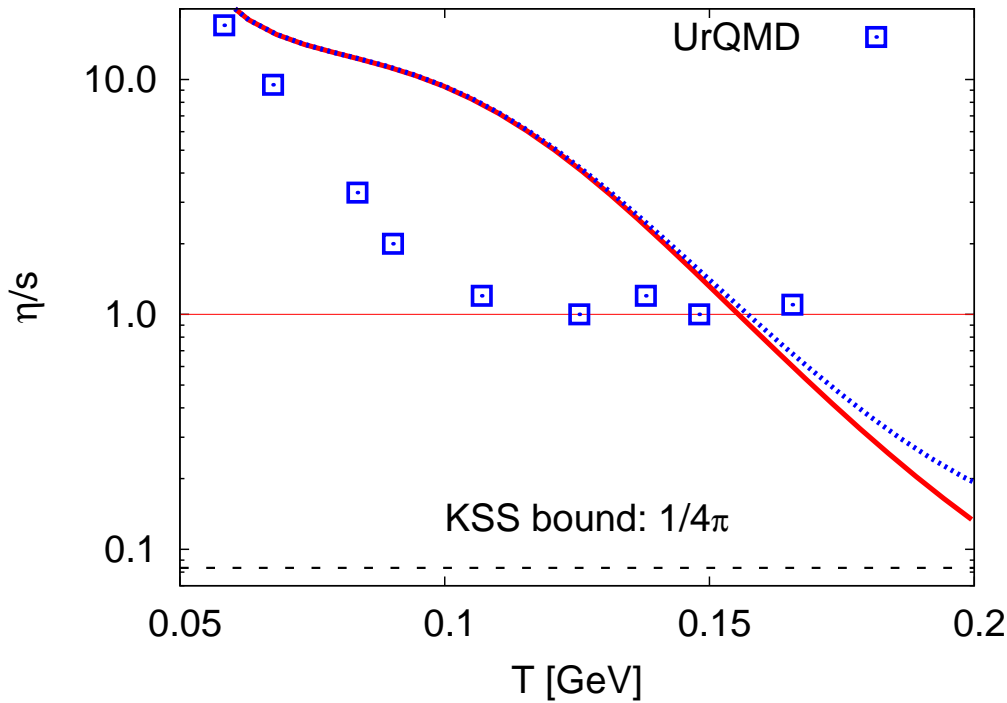}
\caption{The shear viscosity to entropy density ratio $\eta/s$ as
a function of temperature $T$ for the hadron resonance gas with
Hagedorn mass spectrum. Horizontal dashed line give the AdS/CFT lower bound. }
\label{2fig3}
\end{figure}

Using Anti de Sitter space/Conformal Field Theory (AdS/CFT) methods \cite{kss1}, it has been argued that the ratio $\eta/s$ seems to have a universal lower bound in any physical system. The bound is $\eta/s\geq 1/4\pi$. It has been found to this value saturated for a large class of strongly interacting systems with a dual description (string theory in anti-de Sitter space) \cite{class,etas2b}. Also, at temperatures below $T_c$, i.e. Hagerdorn-type models, this ratio has been analyzed using chiral perturbation theory \cite{etas1}, coupled Boltzmann
equations of pions and nucleons in low baryon number densities \cite{etas2}, gas of massless pions \cite{etas2c}, relativistic mean field models with scaled hadron masses and couplings \cite{eta-hadrons4} and viscous relativistic hydrodynamics \cite{etas4}. In Fig. \ref{2fig3}, we draw $\eta/s$ calculated in the present model as a function of $T$. It is clear that $\eta/s$ starts from a much higher values than the AdS/CFT lower bound one. It comes closer to it with increasing $T$. 

Also, it is obvious that $\eta/s$ would be reduced with increasing the Hagedorn mass spectrum. It is a universal property that the quantities which are depending on thermodynamics are suppressed with increasing the mass. Shear viscosity $\eta$ apparently follows this behavior, which can be realized primarily due to the  enhancement of the massive resonances that leads to a decrease in the mean free path of a particle and a corresponding increase in the average binary collision cross section. The results are compatible with the ones introduced in Ref. \cite{eta-hadrons3b} which are compared with the UrQMD simulations. Nevertheless, we notice however, that for Hagedorn resonance gas treated within the rate equation approach, an upper limit of $\eta/s$ is found to
	be as small as the KSS lower bound near $T_c$ \cite{eta-hadrons2}. 

The UrQMD model calculations give saturated $\eta/s$--values in the region $0.125<T>0.166\;$GeV \cite{eta-hadrons3b}. Within this interval, $\eta$ quantitatively equal to the entropy density $s$. Another difference between UrQMD and the present work is graphically shown in the $T$--region, $0.06<T>0.16\;$GeV, where it appears that the UrQMD model drastically underestimates the values of $\eta/s$. It is clear that the massive resonances likely contribute dominantly, especially high $T$. The ratio $\eta/s$ gradually reduces.

The effects of excluded volume on $\eta/s$, which as discussed previously takes into consideration -- at least -- the van der Waals repulsive interactions, are also illustrated graphically in the same figure. The viscosity coefficients are enhanced due to this additional interaction. Finally, we notice that the values of $\eta/s$ very close to $T_c$ are close to the upper bound of $0.24$ obtained from viscous hydrodynamic calculations \cite{upper} of elliptic flow. 

To compare with the lattice QCD calculations below $T_c$ \cite{shin}, we need to rearrange the configurations the Hagedorn fluid to be compilable with the lattice QCD configurations. Since, it is not trivial to match the two configurations, we think that a future study should be devoted to this subject. At higher $T$, the universal properties of bulk viscosity near the QCD phase transition are introduced in Ref. \cite{karsch1}. On the other hand, there are several lattice QCD estimations for the lower bound \cite{nakamuraMeyer}. It is very essential to mention that these lattice QCD calculations have two limitations. First is the use of quench approximation, i.e. without quark pair creation-annihilation effects on vacuum, and second is the use of an ansatz for the spectral function. 

\begin{thebibliography}{99}

\bibitem{larry} M.~Gyulassy and L.~McLerran, Nucl. Phys. A {\bf 750} 30
(2005). 

\bibitem{V-hadron1} M. Ostrowski, Acta Phys. Polon. B {\bf 10}, 875 (1979). 

\bibitem{V-hadron2} A. Hosoya and K. Kajantie, Nucl. Phys. B {\bf 250}, 666 (1985); A. Hosoya, M.-A. Sakagami and M. Takao, Ann, Phys. {\bf 154}, 229 (1984).

\bibitem{redl1} C. Sasaki and K. Redlich, Phys. Rev. C {\bf 79}, 055207 (2009).

\bibitem{etas2} J.-W. Chen, Y.-H. Li, Y.-F. Liu, and E. Nakano, Phys. Rev. D {\bf 76}, 114011 (2007).

\bibitem{hagdr1} R. Hagedorn, Astron. \& Astrophys. {\bf 5}, 184 (1970).

\bibitem{mueller2} D.~Kharzeev and K.~Tuchin, JHEP {\bf 0809}, 093 (2008).

\bibitem{perturQCD1} P. B. Arnold, C. D. Dogan and G. D. Moore, Phys. Rev. D {\bf 74},
  085021 (2006).

\bibitem{karsch1} F. Karsch, D.~Kharzeev and K.~Tuchin, Phys. Lett. B {\bf 663}, 217 (2008). 

\bibitem{QCD-conformal1} G. Torrieri, B. Tomasik and I. Mishustin,
  Phys. Rev. C {\bf 77}, 034903 (2008).

\bibitem{eta-hadrons1} A. Muronga, Phys. Rev. C {\bf 69}, 044901 (2004); 
K. Itakura, O. Morimatsu and H. Otomo, Phys. Rev. D  {\bf 77}, 014014 (2008); 
M. I. Gorenstein, M. Hauer and O. N. Moroz, Phys. Rev. C {\bf 77}, 024911 (2008);


\bibitem{etas2b} J.-W. Chen and E. Nakano, Phys.  Lett. B {\bf 647}, 371 (2007).

\bibitem{eta-hadrons2} J. Noronha-Hostler, J. Noronha and C. Greiner, Phys. Rev. Lett. {\bf 103}, 172302 (2009). 

\bibitem{eta-hadrons3a} A. Muronga, Eur Phys J. ST {\bf 155}, 107 (2008). 

\bibitem{eta-hadrons3b} S. Pal, Phys. Lett. B {\bf 684}, 211 (2010).

\bibitem{eta-hadrons4} A. S. Khvorostukhin, V.D. Toneev and D.N. Voskresensky, arXiv:0912.2191v2 [nucl-th]; arXiv:1003.3531v1 [nucl-th]. 

\bibitem{tawf2} A. Tawfik, Phys. Rev. D {\bf 71}, 054502 (2005); Phys. Lett. B {\bf 623}, 48 (2005); Europhys. Lett. {\bf 75}, 420 (2006); Nucl. Phys. A {\bf 764}, 387 (2006); Fizika B. {\bf 18}, 141 (2009).

\bibitem{hresons} R. Hagedorn, Nuovo Cimento {\bf 35}, 395 (1965).

\bibitem{extndd} J. I. Kapusta and K. A. Olive, Nucl. Phys. A {\bf 408}, 478 (1983).

\bibitem{extndd2} J. Noronha-Hostler, C. Greiner and I. A. Shovkovy, Phys.
Rev. Lett. {\it 100}, 252301 (2008).

\bibitem{tawf1} F. Karsch, K. Redlich and A. Tawfik, Eur. Phys. J. C {\bf 29}, 549 (2003); Phys. Lett. B {\bf 571}, 67 (2003).

\bibitem{reff1} P. Braun-Munzinger, D. Magestro, K. Redlich, and J. Stachel, Phys.
Lett. {\bf B518}, 41 (2001); F. Becattini, J. Cleymans, A. Keranen, E Suhonen and  K. Redlich, Phys. Rev. C {\bf 64}, 024901 (2001).

\bibitem{eckrt} C. Eckart, Phys. Rev. {\bf 58}, 919 (1940).

\bibitem{relaxx} F. Reif, Fundamentals of statistical and Thermal Physics, McGraw-Hill, New York, (1965).

\bibitem{maartns} Roy Maartens, {\it ``Causal thermodynamics in relativity''}, arXiv:astro-ph/9609119.

\bibitem{ref:tau1} M. Prakash, M Prakash, R. Venugopalan and G. Welke, Phys. Reps., {\bf 227}, 321 (1993). 

\bibitem{xiovereta1} S. Gavin, Nucl. Phys. A {\bf 435}, 826 (1985); J.-W. Chen and J. Wang, Phys. Rev. C {\bf 79}, 044913 (2009).

\bibitem{GSt} H. St\"ocker and W. Greiner, Phys. Rep. {\bf 137}, 277 (1986).

\bibitem{etaxiratio1} A. Buchel, Phys. Lett. B 663, 286 (2008); A. Buchel and C. Pagnutti, Nucl. Phys. B {\bf 816}, 62 (2009).

\bibitem{gmat} P. Ring and P. Schuck, The nuclear many-body problem, Springer-Verlag Berlin Heidelberg (1980)

\bibitem{eta-hadrons5} I. Senda, Phys. Lett. B 263, 270 (1991).

\bibitem{newRafls} I. Kuznetsova and J. Rafelski, arXiv:1002.0375 [hep-th]

\bibitem{tawf3} A. Tawfik, AIP Conf. Proc. {\bf 1115}, 239 (2009); arXiv:1002.0296 [gr-qc]; A. Tawfik, M. Wahba, H. Mansour and T. Harko, arXiv:1001.2814 [gr-qc]; Invited talk at 7th International Conference on Modern Problems of Nuclear Physics, Tashkent, Uzbekistan, 22-25 Sep 2009, arXiv:0911.4105 [gr-qc]; A. Tawfik, H. Mansour and M. Wahba, Talk given at 12th Marcel Grossmann Meeting on General Relativity (MG 12), Paris, France, 12-18 Jul 2009, arXiv:0912.0115 [gr-qc].

\bibitem{kss1} P. Kovtun, D. T. Son and A. O. Starinets, JHEP {\bf 0310}, 064 (2003); Phys. Rev. Lett. {bf 94}, 111601 (2005).

\bibitem{class} G. Policastro, D.T. Son, and A.O. Starinets, Phys. Rev. Lett. {\bf 87}, 081601 (2001); JHEP {\bf 0209}, 043 (2002); C.P. Herzog, J. High Energy Phys. {\bf 0212}, 026 (2002); A. Buchel and J.T. Liu, Phys. Rev. Lett. {\bf 93}, 090602 (2004). 

\bibitem{etas1} D. Fernandez-Frailea and A. G. Nicolab, Eur. Phys. J. A {\bf 31}, 848-850 (2007).


\bibitem{etas2c} J.-W. Chen and J. Wang, Phys. Rev. C {\bf 79}, 044913 (2009).

\bibitem{etas4} N. Demir and S. A. Bass, Phys. Rev. Lett. {bf 102}, 172302 (2009).


\bibitem{upper} P. Romatschke and U. Romatschke, Phys. Rev. Lett. {\bf 99}, 172301 (2007); H. Song and U. Heinz, Phys. Lett. B {\bf 658}, 279 (2008).

\bibitem{shin} S. Muroya, talk at  international Workshop on "Hadron Physics and Property of High Baryon Density Matter", Xi'an, China, 22-25 Nov. (2006), hep-ph/0702220

\bibitem{nakamuraMeyer} H.B. Meyer, Phys. Rev. D {\bf 76}, 101701 (2007); 
A. Nakamura and S. Sakai, Nuclear Physics A {\bf 774}, 775-778 (2006). 

\end{thebibliography}
\end{document}